\documentclass[twocolumn,showpacs,amsmath]{revtex4}

\usepackage{graphicx}
\usepackage{dcolumn}
\usepackage{bm}

\newcommand{\be}{\begin{equation}}
\newcommand{\ee}{\end{equation}}
\newcommand{\bea}{\begin{eqnarray}}
\newcommand{\eea}{\end{eqnarray}}
\newcommand{\ba}{\begin{eqnarray}}
\newcommand{\ea}{\end{eqnarray}}

\newcommand{\eqn}[1]{(\ref{#1})}
\newcommand{\beq}{\begin{equation}}
\newcommand{\eeq}{\end{equation}}
\newcommand{\beqa}{\begin{eqnarray}}
\newcommand{\eeqa}{\end{eqnarray}}
\newcommand{\beqar}{\begin{eqnarray*}}
\newcommand{\eeqar}{\end{eqnarray*}}

\newcommand{\reef}[1]{(\ref{#1})}

\newcommand{\eg}{{\it e.g.,}\ }
\newcommand{\ie}{{\it i.e.,}\ }

\def\sac{\, , \,\,\,\,\,}

\def\nc {N_\mt{c}}
\def\nf {N_\mt{f}}

\def\t7 {T_\mt{D7}}

\newcommand{\mq}{M_\mt{q}}      

\newcommand{\vep}{\varepsilon}

\newcommand{\mt}[1]{\textrm{\tiny #1}}

\begin{document}

\preprint{hep-th/0610184}

\title{Holographic Viscosity of Fundamental Matter} 

\author{David Mateos,$^1$ Robert C. Myers,$^{2,3,4}$ and
Rowan M. Thomson,$^{3,4}$} \affiliation{$^1\,$Department of Physics,
University of California, Santa
Barbara, CA 93106-9530, USA \\
$^2\,$Kavli Institute for Theoretical Physics, University of
California, Santa Barbara, CA 93106-4030, USA \\
$^3 \,$Perimeter Institute for Theoretical Physics, Waterloo,
Ontario N2L 2Y5, Canada \\
$^4 \,$Department of Physics and Astronomy, University of Waterloo,
Waterloo, Ontario N2L 3G1, Canada }


\begin{abstract}
A holographic dual of a finite-temperature $SU(\nc)$ gauge theory
with a small number of flavours $\nf \ll \nc$ typically contains
D-branes in a black hole background. By considering the backreaction
of the branes, we demonstrate that, to leading order in $\nf/\nc$,
the viscosity to entropy ratio in these theories saturates the
conjectured universal bound $\eta/s \geq 1/4\pi$. The contribution
of the fundamental matter $\eta_\mt{fund}$ is therefore enhanced at
strong 't Hooft coupling $\lambda$; for example, $\eta_\mt{fund}
\sim \lambda \nc \nf T^3$ in four dimensions. Other transport
coefficients are analogously enhanced. These results hold with or
without a baryon number chemical potential.
\end{abstract}

\maketitle

\noindent {\bf Introduction:} A universal bound was recently
proposed \cite{KSS} for the ratio of the shear viscosity to the
entropy density of any physical system as $\eta/s\ge1/4\pi$. In
particular, this bound is conjectured to hold for all relativistic
quantum field theories at finite temperature that exhibit
hydrodynamic behaviour at long wave-lengths. Perhaps surprisingly,
experimental results from the Relativistic Heavy Ion Collider (RHIC)
suggest that, for QCD just above the deconfinement phase transition,
the value of $\eta/s$ is close to saturating this bound
\cite{shuryak}. This would indicate that the quark-gluon plasma
formed at RHIC is an almost perfect liquid.

Unfortunately, at present there are no theoretical tools with which
to calculate transport coefficients in QCD in this strong coupling
regime, \eg the viscosity. However, a large class of gauge theories
are accessible to study with the gauge/gravity correspondence
\cite{maldacena}. In particular, in the gauge theory limit of large
$\nc$ and large 't Hooft coupling $\lambda$, the dual description
reduces to classical supergravity. In fact, the proposal \cite{KSS}
for a universal bound on $\eta/s$ originated with calculations in
this holographic context. Explicit calculations
\cite{PSS1,PSS2,explicit1,explicit2} and general arguments
\cite{KSS,general,general2} have demonstrated that the bound is
exactly saturated by a large class of holographic theories in the
limit cited above. In order to make contact with real-world QCD, it
is clearly important to consider $1/\lambda$ and $1/\nc$
corrections. For four-dimensional ${\cal N} =4$ super Yang-Mills
(SYM), the leading correction of the first type was shown to raise
the value of $\eta/s$ above the bound \cite{lambda}. That is, at
finite coupling the bound still holds but is no longer saturated.

A feature common to all of the gauge theories considered in these
hydrodynamic studies is that the matter degrees of freedom transform
in the adjoint representation of the gauge group \cite{exception}.
In this letter, we study the effect of adding matter fields
transforming in the fundamental representation, bringing us one step
closer to QCD. In particular, we focus on four-dimensional $SU(\nc)$
SYM coupled to $\nf$ fundamental hypermultiplets with $\nf \ll \nc$.
Large-$\nc$ counting rules imply that, in the deconfined phase, the
contribution of the gluons and adjoint matter to physical quantities
is of order $\nc^2$. Further, the first correction in the absence of
fundamental matter is of order 1, \ie the relative contribution is
suppressed by $1/\nc^2$. Instead, the relative contribution of
fundamental matter is only suppressed by $\nf/\nc$, and therefore it
constitutes the leading correction in the large-$\nc$ limit.

The dual gravity description is given by $\nf$ D7-brane probes
\cite{flavour} in the background geometry of $\nc$ D3-branes. At
finite temperature, the latter contains a black hole \cite{witten}.
We will show that, to leading order in $\nf/\nc$, the calculation of
the $\eta/s$ ratio can be effectively reduced to one in
five-dimensional Einstein gravity coupled to a scalar field. General
results \cite{KSS,general} then guarantee that $\eta/s=1/4\pi$.
Since the D7-brane contribution to the entropy density is known to
be of order $s_\mt{fund} \sim \lambda \nc \nf T^3$ \cite{PRL}, this
implies that the contribution of the fundamental matter to the shear
viscosity at strong 't Hooft coupling is enhanced with respect to
that dictated solely by large-$\nc$ counting rules.

In the final section we will argue that an analogous enhancement
takes place for other transport coefficients. We will also explain
how our results extend straightforwardly to other holographic gauge
theories described by Dp/Dq \cite{d4d6,PRL} or
Dp/Dq/D$\bar{\mbox{q}}$ \cite{d8d8bar} systems, as well as to
systems with a non-zero baryon number chemical potential.
We will finish with some comments on effects beyond order $\nf/\nc$.

\noindent {\bf Holographic Framework:} The shear viscosity of the
gauge theory in a two-plane labelled by $x^i,x^j$ may be extracted
from the retarded correlator of  two stress energy tensors via
Kubo's formula
\beq \eta = \lim_{\omega \to 0} \frac{1}{2\omega} \int dt\, d^3x \,
e^{i \omega t} \left< [T_{ij}(x),T_{ij}(0)] \right> \,, \label{kubo}
\eeq
where no summation over $i,j$ is implied. The stress energy tensor
is dual on the string side to a metric perturbation $H_{ij}$
polarised along the same two-plane. The two-point function above may
be calculated by taking two functional derivatives of the on-shell
string effective action with respect to this perturbation
\cite{recipe}. In the large-$\nc$, large-$\lambda$ limit, this
effective action reduces to the  type IIB supergravity action
coupled to the worldvolume action of the D7-branes, $I = I_\mt{IIB}
+ I_\mt{D7}$. Schematically, we have:
\be I = \frac{1}{16 \pi G} \int d^{10} x \sqrt{-g} R - \nf T_\mt{D7}
\int d^8 x \sqrt{-g_\mt{ind}} + \cdots \,, \label{action} \ee
where $g_\mt{ind}$ is the induced metric on the D7-branes. In terms
of the string length and coupling:
\be 16\pi G=(2\pi)^7 \ell_s^8 g_s^2 \sac T_\mt{D7}=2\pi/(2\pi
\ell_s)^8 g_s \,. \ee
The ratio between the normalisations of the two terms above is
\be \vep = 16\pi G \nf \t7  = \frac{\lambda}{2\pi}\,\frac{\nf}{\nc}
\,, \ee
where $\lambda = g^2_\mt{YM} \nc = 2\pi g_s \nc$ is the 't Hooft
coupling. This ratio controls the relative magnitude of the
D7-branes' contribution to physical quantities, \eg the entropy
density \cite{PRL}. We will assume that $\vep \ll 1$ and hence that
the D7-branes can be treated as a small perturbation; for fixed
$\lambda$ this is achieved by taking $\nf\ll\nc$. We will begin by
examining contributions of order $\vep$ in the next section. In the
last section, we will comment on effects of order $\vep^2$ and
higher.

In the absence of D7-branes, the supergravity background dual to
four-dimensional ${\cal N}=4$ SYM at temperature $T$ is (in the
notation of \cite{PRL})
\bea
ds^2 &=&  ds^2_\mt{5} + L^2 d\Omega_{5}^2 \,, \label{D3geom} \\
ds^2_\mt{5} &=& \frac{(\pi L T \rho)^2}{2} \left[-\frac{f^2}{\tilde
f}dt^2 + \tilde{f} dx^2_i \right] + \frac{L^2}{\rho^2} d\rho^2  \,,
\label{D3geom5} \eea
where
\be f(\rho)=1-1/\rho^4 \sac \tilde{f}(\rho)=1+1/\rho^4 \,, \ee
and $L=(4\pi g_s \nc)^{1/4} \ell_s$ is the asymptotic AdS radius.
There are also $\nc$ units of Ramond-Ramond (RR) flux through the
five-sphere while the remaining supergravity fields vanish. The
metric (\ref{D3geom},\ref{D3geom5}) possesses an event horizon at
$\rho=1$. The entropy density of the gauge theory is then given by
the geometric entropy of the horizon \cite{3/4}
\be s = \frac{\pi^3}{4G_\mt{5}} L^3 T^3 =
 \frac{\pi^2}{2}\nc^2\,T^3 \,, \label{entropy} \ee
where $G_\mt{5} = G/\pi^3 L^5$ is the five-dimensional Newton's
constant obtained by dimensional reduction on the five-sphere.

Now we introduce the D7-branes oriented such that five worldvolume
directions match those of the five-dimensional black hole
\eqn{D3geom5}, $y \equiv \{ t,x^i,\rho \}$, and the remainder wrap
an $S^3$ (with a possibly varying radius) inside the $S^5$ of
\eqn{D3geom}. We adapt coordinates in this internal space such that
\beqa
d\Omega^2_5 =  d\theta^2+\sin^2\theta d\Omega^2_3+
\cos^2\theta d\phi^2  \label{met2}
\eeqa
and describe the D7-branes embedding as $\chi=\chi(\rho)$, with
$\chi=\cos \theta$. To order $\vep^0$, this is determined by
extremising the D7-brane action in the background
(\ref{D3geom},\ref{D3geom5}). Asymptotically, one finds
\be
\chi = \frac{m}{\rho} + \frac{c}{\rho^3} + \cdots \,,
\ee
where $m$ and $c$ are proportional to
the quark mass $\mq$ and condensate
$\langle \bar{\psi} \psi \rangle$, respectively. In the interior,
the D7-branes may or may not reach the black hole horizon
\cite{PRL}.

\noindent {\bf Viscous Branes:} As alluded to above, the calculation
of the shear viscosity proceeds as follows. First, one solves the
(linearised) equation of motion for a metric perturbation $H$ around
the appropriate background. Next, one evaluates the appropriate
action for the perturbed background to quadratic order in $H$. A
second derivative of the on-shell action then yields the desired
two-point function \cite{recipe} with which the Kubo formula
\eqn{kubo} is evaluated.

In the absence of D7-branes the appropriate action is $I_\mt{IIB}$
and the background is given by (\ref{D3geom}). In the presence of
the D7-branes the relevant action is instead that in
eq.~\eqn{action}. To first order in the $\vep$-expansion, this
affects the calculation of the viscosity in three ways. First, the
branes will produce ${\cal O}(\vep)$ corrections to the metric
\eqn{D3geom}, as well as to the dilaton and the RR axion, since they
act as new sources in the field equations arising from the combined
action. These background corrections then lead to modifications of
the field equation satisfied by $H$. Second, the branes will also
modify the $H$ field equation directly through the extra source
terms originating from the variation of $I_\mt{D7}$ with respect to
$H$. Third, the second-derivative of the on-shell action, which
yields the correlator in Kubo's formula, may acquire contributions
from $I_\mt{D7}$. \cite{higher}

We will now show that only the first two types of possible modifications
do actually contribute and, moreover, that the only type of
background correction that needs to be considered is the
zero-mode (on the five-sphere) of the five-dimensional black hole
metric \eqn{D3geom5}.

We begin by considering the third set of possible contributions
listed above. Expanding the brane action around the ${\cal
O}(\vep^0)$ background, one finds that, because $H$ enters the
action non-derivatively, the $H^2$ terms do not have a form which
will contribute in the Kubo formula \cite{recipe}. However, turning
on $H$ also induces a correction $\delta \chi = {\cal O} (H^2)$ in
the embedding of the branes. This leads to a surface term in the
variation of the D7-brane action,
\be \delta I_\mt{D7} \sim \left. \frac{\partial {\cal
L}_\mt{D7}}{\partial (\partial_\rho \chi)} \delta \chi
\right|_{\rho_\mt{max}} \,,
\ee
of the right form to contribute to the two-point correlator.
However, this contribution is proportional to $\delta m$ \cite{PRL}
and hence vanishes because the variation of the action with respect
to $H$ must be taken while keeping the quark mass fixed.

Consider now corrections to the background \eqn{D3geom}.
While there are ${\cal O}(\vep)$ corrections to the dilaton and the RR axion,
these only produce ${\cal O}(\vep^2)$ contributions to the $H$ field equation
because they enter the supergravity action quadratically (in the Einstein
frame). We are thus left to consider the contributions of
corrections to the spacetime metric. Schematically, to order $\vep$,
the background metric takes the form
\be g = g_0(y) + \vep g_1^{(0)}(y)+\vep \sum_{\ell \neq 0}
g_1^{(\ell)}(y) Y^{(\ell)}(\Omega) \,, \ee
where $Y^{(\ell)}(\Omega)$ are spherical harmonics on $S^5$.  The
zero'th order metric \eqn{D3geom} and the correction $g_1^{(0)}$ are
constant on the $S^5$. However, since the D7-branes only fill an
$S^3$ in the internal space, they also source the $g_1^{(\ell)}$
corrections which vary over the $S^5$. For the following, an
important point is that the functions $g_1^{(0)}(y)$ and
$g_1^{(\ell)}(y)$ respect the symmetries of the background geometry
\eqn{D3geom} and the brane embedding, \ie translations in
$\{t,x^i\}$ and $SO(3)$ rotations in $x^i$, as well as $SO(4)$
rotations in the internal $S^3$ wrapped by the D7-brane. The
perturbation $H$ has a similar decomposition:
\be H = H_0(y) + \vep H_1^{(0)}(y)+\vep \sum_{\ell \neq 0}
H_1^{(\ell)}(y) Y^{(\ell)}(\Omega) \,. \ee
In the absence of D7-branes, one works consistently with the
$S^5$-independent perturbation $H_0$ \cite{PSS1,PSS2}. However, in the
presence of the D7-branes, nontrivial $H_1^{(\ell)}$ are sourced
when $H_0$ is turned on. Indeed, after integration over the $S^5$,
the supergravity action produces couplings of the schematic form
$\vep^2 \int d^5y\, H_0 H_1^{(\ell)} g_1^{(\ell)}$. Similarly, the
D7-branes action produces couplings like $\vep^2 \int d^5y\, H_0
H_1^{(\ell)}$, which arise from spherical harmonics $Y^{(\ell)}$
that are constant on the $S^3$ wrapped by the D7-branes. However, as
indicated, both types of terms are of order $\vep^2$ and so we may
neglect their contribution since here we only wish to determine the
correlator \eqn{kubo} up to order $\vep$.

We therefore conclude that, to order $\vep$, we need only consider
the zero-modes $g_1^{(0)}(y)$ and $H_1^{(0)}(y)$, and may drop all
modes with non-trivial dependence on the $S^5$ directions. Hence in
working to order $\vep$, evaluation of the viscosity actually
reduces to a five-dimensional calculation. We can make the latter
concrete by dimensionally reducing the action \eqn{action} to five
dimensions ignoring all the Kaluza-Klein modes on the five-sphere,
as well as the other supergravity fields. By the previous arguments,
the resulting action still captures all of the relevant fields for
the calculation of the viscosity to this order. The five-dimensional
action can be written as
\bea I_\mt{5} &=& \frac{1}{16 \pi G_\mt{5}} \int d^{5} y \sqrt{-g}
\left[R +\frac{12}{L^2}- \frac{2\,\vep}{\pi L^2}(1-\chi^2)
\right. \nonumber\\
&& \qquad\qquad\quad\left.\times\ \sqrt{1-\chi^2 + L^2 g^{\rho \rho}
(\partial_\rho \chi)^2}\ \right] \,,\label{action5}  \eea
where $g$ denotes the five-dimensional metric. The first two terms
originate from the reduction of $I_\mt{IIB}$, whereas the last one
comes from the reduction of $I_\mt{D7}$ \cite{PRL}. Note that, in
the action \reef{action5}, we have  only allowed scalar field
configurations depending on the radial coordinate, since this
suffices for our purposes. This system is just five-dimensional
Einstein gravity coupled to a cosmological constant and a(n unusual)
scalar field $\chi$. In an $\vep$-expansion, the black hole
solutions generated by this auxiliary theory will match the
asymptotically AdS part of the original ten-dimensional solution to
order $\vep$, \ie the brane profile $\chi(\rho)$ and the background
metric \eqn{D3geom5} plus the order-$\vep$ correction
$g_1^{(0)}(\rho)$.

The viscosity may now be obtained by calculating the perturbation
$H_{ij}$ around the five-dimensional solution and taking the second
functional derivative of the action \eqn{action5} evaluated
on-shell. However, the black hole solutions of our auxiliary
five-dimensional system satisfy the symmetries required in
\cite{KSS}, and hence the result is guaranteed to satisfy
$\eta/s=1/4\pi$. We thus conclude that this universal bound is still
saturated in the full ten-dimensional string theory when working to
first order in $\vep$. An immediate consequence is that the
contributions of the fundamental matter to the viscosity and the
entropy density are related (within our approximations) as
$\eta_\mt{fund} = s_\mt{fund}/4\pi$. The leading contribution to the
entropy density was determined \cite{PRL} to be
\be s_\mt{fund} = \frac{\lambda}{16} \nc \nf\, T^3\ h\! \left(
\frac{\lambda T}{\mq} \right) \,, \ee
where the function $h(x)$ satisfies $h(0)=0$, $h(\infty)=1$,  and
makes a cross-over between both values around $x\sim1$. Note that
this cross-over includes a small discontinuity arising from a
first-order phase transition of the fundamental matter \cite{PRL}.
We therefore conclude that both $s_\mt{fund}$ and $\eta_\mt{fund}$
are enhanced at strong 't Hooft coupling with respect to the ${\cal
O}(\nc \nf)$-value dictated solely by large-$\nc$ counting rules.

The calculation of $s_\mt{fund}$ in \cite{PRL} was performed by
identifying the Euclidean action of the D7-branes with
$F_\mt{fund}/T$, where $F_\mt{fund}$ is the free energy contribution
of the fundamental matter. The entropy is then determined as
$s_\mt{fund} = -\partial F_\mt{fund} / \partial T$. This entropy
should, of course, coincide with the change in the horizon area
induced by the presence of the D7-branes. The latter can be
explicitly verified for the case of massless quarks, which
corresponds to an `equatorial embedding' with $\chi=0$. In this
case, the result from \cite{PRL} for the entropy density is
$s_\mt{fund} = \lambda \nc \nf \,T^3/16$. In the action
\eqn{action5}, we see that the net effect of these `equatorial'
D7-branes is to shift the effective cosmological constant. The
corresponding black hole solution is still given by \eqn{D3geom5},
with the replacement $L^2 \rightarrow L^2 /(1-\vep/6\pi)$. The same
replacement in \eqn{entropy} shifts the entropy to order $\vep$ by
$\delta s
= \lambda \nc \nf \,T^3/16$, in perfect agreement with the previous
result.

\noindent {\bf Discussion:} We have seen that the calculation of the
contribution of fundamental matter to the shear viscosity may be
effectively reduced to a calculation in five dimensions. An
analogous simplification takes place for other transport
coefficients that can be extracted from correlators involving local
operators with vanishing R-charge, since these are dual to modes
that carry no angular momentum on the $S^5$. Examples involving
components of the stress-energy tensor include the speed of sound
$v_s$ and the bulk viscosity $\xi$. Other transport coefficients
that involve R-charged operators, such as the R-charge diffusion
constant \cite{PSS2}, or extended strings, such as the jet quenching
parameter $\hat{q}$ (see \eg \cite{qhat}), may require a ten-dimensional
calculation. Generically, however, we expect the relative
contribution of the fundamental matter to be of order $\vep \sim
\lambda \nf / \nc$, since this controls the backreaction of the
branes. A special case is the bulk viscosity, which may be extracted
from a two-point function of $T_{ii} -  v_s^2 T_{00}$ \cite{bulk}.
This combination vanishes if $\nf=0$ by conformal invariance, and
hence $\xi = {\cal O} (\vep^2)$ in the presence of fundamental
matter.

Above, our discussion focussed on the D3/D7 system, but the
arguments are easily extended to a more general system of Dq-branes
in a Dp-background. In particular, systems arising from Dp- and
Dq-branes intersecting over $d$ common spatial directions have been
of some interest \cite{d4d6,generalpq}. These constructions are dual
to a finite-temperature SYM theory in $p+1$ dimensions coupled to
fundamental matter confined to a $(d+1)$-dimensional defect. One new
feature in these generalised configurations is that the defect
breaks translational invariance along the $p-d$ orthogonal
directions. In order to calculate the shear viscosity \cite{corrx}
along the translationally invariant directions parallel to the
defect, the simplest approach is to compactify these extra
directions \cite{vol}. The arguments in the previous section go
through essentially unchanged except for the fact that the index
$\ell$ now labels momentum modes both along the $S^{8-p}$ transverse
to the Dp-branes and along the $p-d$ directions orthogonal to the
defect. In this case the problem of calculating the leading
contribution of the fundamental matter to the viscosity/entropy
ratio can be reduced to a calculation in $(d+2)$-dimensional
Einstein gravity coupled to a set of scalar fields. In addition to
the scalar $\chi$ above describing the size of the internal $S^n$ of
the Dq-brane, this set now includes the dilaton and the metric
components governing the size of the internal $S^{8-p}$ of the
background geometry and the size of the $(p-d)$-dimensional space
orthogonal to the defect \cite{wow}. This lower dimensional theory
again captures all of the relevant fields to calculate the viscosity
to leading order in $\nf/\nc$. Further, the form of the
$(d+2)$-dimensional gravity theory and the background guarantees
that $\eta/s=1/4\pi$. The leading result for the entropy density was
determined in \cite{PRL} and hence we have
\be \eta_\mt{fund} \sim \nc \nf\, T^d\,
g_\mt{eff}(T)^{\frac{2(d-1)}{5-p}} \,, \ee
where $g^2_\mt{eff}(T) = \lambda T^{p-3}$ is the dimensionless
effective 't Hooft coupling for a $(p+1)$-dimensional theory at
temperature $T$ \cite{itz}. Here the gauge/gravity duality is only
valid in the strongly coupled regime \cite{itz} and hence we see
again an enhancement beyond the large-$\nc$ counting. It would be
interesting to understand if this enhancement extends to other
transport properties in the same way, as was found for the D3/D7
case. The same line of argument can also be implemented for the
Dp/Dq/D$\bar{\mbox{q}}$ systems which have also been studied
recently \cite{d8d8bar}.

The $U(\nf) \simeq SU(\nf) \times U(1)_\mt{B}$ gauge symmetry on the
Dq-branes is a global, flavour symmetry of the dual gauge theory.
The results above also hold when a baryon number chemical potential
for the $U(1)_\mt{B}$ charge is introduced. This is dual
\cite{chemical} to turning on the time component of the gauge
potential on the Dq-branes, $A_0(\rho)$ \cite{Rcharge}. The
arguments of the previous section again go through essentially
unchanged, except for the fact that an additional vector $A_\mu$ is
added to the reduced $(d+2)$-dimensional Einstein gravity theory.
Thus the saturation of the bound is not affected by the introduction
of a chemical potential.

Above we have worked to the lowest order in the parameter $\vep \sim
\lambda \nf/\nc$ that controls the backreaction of the D7-branes on
the D3-brane geometry. We have argued that to this order one may
ignore all effects of this backreaction except for those on the
non-compact part of the metric. We regard the agreement between the
entropy density as calculated in \cite{PRL} and as obtained here
from the change in the horizon area as a quantitative consistency
check of this approach. In calculating beyond order $\vep$, the
internal modes, \eg $g_1^{(\ell)}(y)$, and other supergravity
fields, \eg the RR axion, will all play a role. Further at ${\cal
O}(\vep^2)$, quantum effects will have to be considered
\cite{hawking}. Closed string loop corrections naturally appear in
an expansion in $g_s^2 \sim \lambda^2/\nc^2$. Thus if as above $\nf$
is fixed, the loop corrections may be of the same magnitude as the
higher order D7-brane contributions, \ie $g_s^2 \sim \vep^2$ if $\nf
= {\cal O}(1)$.
 
In order to suppress string loop corrections with respect to the
backreaction of the D7-branes (or more generally the Dq-branes), one
must keep $\nf / \nc$ fixed while taking $\nc \rightarrow \infty$.
In this limit, the entire backreaction of the branes must be taken
into account. A fully backreacted solution is singular for the D3/D7
system \cite{sing}, since the dual gauge theory possesses a Landau
pole at some finite scale $\Lambda_\mt{UV}$. This of course should
not affect the hydrodynamic behaviour as long as the other scales in
the problem, the temperature and the quark mass, are much lower than
$\Lambda_\mt{UV}$  (just like the transport properties of an
electromagnetic plasma are not affected by the Landau pole in QED).
Similarly, in the case of a general Dp-brane background (with
$p\ne3$), one finds strong coupling or curvature divergences in the
far UV, which are irrelevant for the long-wavelength hydrodynamics.
In these cases, the classical backreacted solution would
effectively resum the effects of the fundamental matter to leading
order in the $1/\lambda$, $1/\nc$ expansions. It may be possible to
prove that the viscosity bound is still saturated by these solutions
by extending the arguments of \cite{general,general2}. The
finite-temperature backgrounds of \cite{exception1,spanish}
represent a step towards this goal.

\noindent {\bf Acknowledgments:} We thank H. Elvang, S. Mathur, S.
Shenker, and very especially R. Emparan for useful discussions. We also
thank A. Starinets for his collaboration at an early stage. DM
thanks the Aspen Center for Physics and the KITP for hospitality
during the various stages of this project. Research at the Perimeter
Institute is supported in part by funds from NSERC of Canada and
MEDT of Ontario. We also acknowledge support from NSF grant
PHY-0244764 (DM), NSERC Discovery grant (RCM) and NSERC Canada
Graduate Scholarship (RMT). Research at the KITP was supported in
part by the NSF under Grant No. PHY99-07949.


\begin{thebibliography}{99}

\bibitem{KSS}
P.~Kovtun, D.T.~Son and A.O.~Starinets,
  JHEP {\bf 0310}, 064 (2003)
  [arXiv:hep-th/0309213];
  Phys.\ Rev.\ Lett.\  {\bf 94}, 111601 (2005)
  [arXiv:hep-th/0405231];

\bibitem{shuryak}
  E.~Shuryak,
  Prog.\ Part.\ Nucl.\ Phys.\  {\bf 53}, 273 (2004)
  [arXiv:hep-ph/0312227];
  D.~Teaney,
   Phys.\ Rev.\ C {\bf 68}, 034913 (2003)
   [arXiv:nucl-th/0301099].

\bibitem{maldacena}
  J.M.~Maldacena,
  Adv.\ Theor.\ Math.\ Phys.\  {\bf 2}, 231 (1998)
  [Int.\ J.\ Theor.\ Phys.\  {\bf 38}, 1113 (1999)]
  [arXiv:hep-th/9711200].

\bibitem{PSS1}
  G.~Policastro, D.T.~Son and A.O.~Starinets,
  Phys.\ Rev.\ Lett.\  {\bf 87}, 081601 (2001)
  [arXiv:hep-th/0104066].

\bibitem{PSS2}
  G.~Policastro, D.T.~Son and A.O.~Starinets,
  JHEP {\bf 0209} (2002) 043
  [arXiv:hep-th/0205052].

\bibitem{explicit1}
C.P.~Herzog,
  JHEP {\bf 0212}, 026 (2002)
  [arXiv:hep-th/0210126];
A.~Buchel,
  Nucl.\ Phys.\ B {\bf 708}, 451 (2005)
  [arXiv:hep-th/0406200];
P.~Benincasa and A.~Buchel,
  Phys.\ Lett.\ B {\bf 640}, 108 (2006)
  [arXiv:hep-th/0605076].

\bibitem{explicit2}
J.~Mas,
  JHEP {\bf 0603}, 016 (2006)
  [arXiv:hep-th/0601144];
D.T.~Son and A.O.~Starinets,
  JHEP {\bf 0603}, 052 (2006)
  [arXiv:hep-th/0601157];
O.~Saremi,
  arXiv:hep-th/0601159;
K.~Maeda, M.~Natsuume and T.~Okamura,
  Phys.\ Rev.\ D {\bf 73}, 066013 (2006)
  [arXiv:hep-th/0602010];
A.~Buchel and J.T.~Liu,
  arXiv:hep-th/0608002.

\bibitem{general}
A.~Buchel and J.T.~Liu,
  Phys.\ Rev.\ Lett.\  {\bf 93}, 090602 (2004)
  [arXiv:hep-th/0311175];
A.~Buchel,
  Phys.\ Lett.\ B {\bf 609}, 392 (2005)
  [arXiv:hep-th/0408095].

\bibitem{general2}
P.~Benincasa, A.~Buchel and R.~Naryshkin,
  arXiv:hep-th/0610145.

\bibitem{lambda}
  A.~Buchel, J.T.~Liu and A.O.~Starinets,
  Nucl.\ Phys.\ B {\bf 707}, 56 (2005)
  [arXiv:hep-th/0406264];
  P.~Benincasa and A.~Buchel,
  JHEP {\bf 0601}, 103 (2006)
  [arXiv:hep-th/0510041].

\bibitem{exception}
An exception is ref. \cite{exception1}, where fundamental matter was
considered. However, the construction involves D5-branes and so the
temperature is not a free parameter.

\bibitem{exception1}
R.~Casero, C.~Nunez and A.~Paredes,
  Phys.\ Rev.\ D {\bf 73}, 086005 (2006)
  [arXiv:hep-th/0602027].

\bibitem{flavour}
 A.~Karch and L.~Randall,
 JHEP {\bf 0106}, 063 (2001)
  [arXiv:hep-th/0105132];
  A.~Karch and E.~Katz,
  JHEP {\bf 0206}, 043 (2002)
  [arXiv:hep-th/0205236].

\bibitem{witten}
  E.~Witten,
  Adv.\ Theor.\ Math.\ Phys.\  {\bf 2}, 505 (1998)
  [arXiv:hep-th/9803131].

\bibitem{PRL}
  D.~Mateos, R.C.~Myers and R.M.~Thomson,
  Phys.\ Rev.\ Lett.\ {\bf 97}, 091601 (2006)
  [arXiv:hep-th/0605046];
%
   ``Thermodynamics of the brane", to appear.


\bibitem{d4d6}
  M.~Kruczenski, D.~Mateos, R.C.~Myers and D.J.~Winters,
  JHEP {\bf 0405}, 041 (2004)
  [arXiv:hep-th/0311270].

\bibitem{d8d8bar}
T.~Sakai and S.~Sugimoto,
  Prog.\ Theor.\ Phys.\  {\bf 113}, 843 (2005)
  [arXiv:hep-th/0412141];
  {\bf 114}, 1083 (2006)
  [arXiv:hep-th/0507073];
O.~Aharony, J.~Sonnenschein and S.~Yankielowicz,
  arXiv:hep-th/0604161;
N.~Horigome and Y.~Tanii,
  arXiv:hep-th/0608198;
E.~Antonyan, J.A.~Harvey and D.~Kutasov,
  arXiv:hep-th/0608177;
  arXiv:hep-th/0608149;
  arXiv:hep-th/0604017;
A.~Parnachev and D.A.~Sahakyan,
  Phys.\ Rev.\ Lett.\  {\bf 97}, 111601 (2006)
  [arXiv:hep-th/0604173];
Y.~Gao, W.~Xu and D.~Zeng,
  JHEP {\bf 0608}, 018 (2006)
  [arXiv:hep-th/0605138];
D.~Gepner and S.~Sekahr Pal,
  arXiv:hep-th/0608229;
A.~Basu and A.~Maharana,
  arXiv:hep-th/0610087.

\bibitem{recipe}
D.T.~Son and A.O.~Starinets,
  JHEP {\bf 0209}, 042 (2002)
  [arXiv:hep-th/0205051];
C.P.~Herzog and D.T.~Son,
  JHEP {\bf 0303}, 046 (2003)
  [arXiv:hep-th/0212072].

\bibitem{3/4}
  S.S.~Gubser, I.R.~Klebanov and A.W.~Peet,
  Phys.\ Rev.\ D {\bf 54}, 3915 (1996)
  [arXiv:hep-th/9602135].

\bibitem{higher}
At higher orders there are of course other contributions. For
example, at ${\cal O} (\vep^2)$ there is a contribution due to the fact that
background corrections modify the D7-brane embedding.


\bibitem{qhat}
  C.P.~Herzog, A.~Karch, P.~Kovtun, C.~Kozcaz and L.G.~Yaffe,
  JHEP {\bf 0607}, 013 (2006)
  [arXiv:hep-th/0605158];
  H.~Liu, K.~Rajagopal and U.A.~Wiedemann,
  arXiv:hep-ph/0605178;
  S.~S.~Gubser,
  arXiv:hep-th/0605182;
  J.~Casalderrey-Solana and D.~Teaney,
  arXiv:hep-ph/0605199.

\bibitem{bulk}
  A.~Hosoya, M.~Sakagami and M.~Takao,
  Annals Phys.\  {\bf 154}, 229 (1984).

\bibitem{generalpq} D.~Arean and A.V.~Ramallo,
  JHEP {\bf 0604}, 037 (2006)
  [arXiv:hep-th/0602174];
R.C.~Myers and R.M.~Thomson, JHEP {\bf 0609}, 066 (2006)
[arXiv:hep-th/0605017];
D.~Arean, A.V.~Ramallo and D.~Rodriguez-Gomez,
  arXiv:hep-th/0609010.

\bibitem{corrx} The framework for the calculation of
correlators is less well developed for such backgrounds, which are
not asymptotically AdS. Cascading gauge theories are an interesting
case where these techniques are being developed \cite{house}.

\bibitem{house} M.~Krasnitz,
  arXiv:hep-th/0011179;
  JHEP {\bf 0212}, 048 (2002)
  [arXiv:hep-th/0209163];
O.~Aharony, A.~Buchel and A.~Yarom,
  Phys.\ Rev.\ D {\bf 72}, 066003 (2005)
  [arXiv:hep-th/0506002];
  arXiv:hep-th/0608209;
M.~Berg, M.~Haack and W.~Muck,
  Nucl.\ Phys.\ B {\bf 736}, 82 (2006)
  [arXiv:hep-th/0507285];
A.~Buchel,
  Phys.\ Rev.\ D {\bf 72}, 106002 (2005)
  [arXiv:hep-th/0509083].


\bibitem{vol} Compactifying is also required to make
the $(d+2)$-dimensional Newton's constant finite. Since Newton's
constant cancels in the ratio $\eta/s$, there is no obstruction to
taking the infinite volume limit at the end.

\bibitem{wow} One might similarly worry about the RR field sourced by
the Dq-branes. This will be of order $\vep$ and so naturally
contributes to the stress tensor at order $\vep^2$. However, there
may also be order-$\vep$ cross-terms if this RR field already
appears as part of the background generated by the Dp-brane. For
supersymmetric cases \cite{generalpq}, this only happens for the
D3/D3, D2/D4 and D1/D5 systems. In all of these, the defect is
(1+1)-dimensional and so the shear viscosity is not defined.

\bibitem{itz}
N.~Itzhaki, J.M.~Maldacena, J.~Sonnenschein and S.~Yankielowicz,
  Phys.\ Rev.\ D {\bf 58}, 046004 (1998)
  [arXiv:hep-th/9802042].

\bibitem{chemical}
N.~Horigome and Y.~Tanii,
  arXiv:hep-th/0608198;
S.~Kobayashi, D.~Mateos, S.~Matsuura, R.C.~Myers and R.M.~Thomson,
to appear.

\bibitem{Rcharge}
This baryon number chemical potential should not be confused with
the R-charge chemical potential considered in {\it e.g.}
\cite{explicit2,general2}, which is dual in the string description
to angular momentum on the sphere.

\bibitem{hawking}
We expect Hawking radiation contributes at order $\nf^2/\nc^2$. In
the presence of $\nf$ D7-branes, there are (at least) ${\cal
O}(\nf^2)$ species or degrees of freedom into which the black hole
can Hawking radiate. However, this is still suppressed by
$1/\lambda^2$ relative to $\vep^2$ at strong 't Hooft coupling.

\bibitem{sing}
I.~Kirsch and D.~Vaman,
  Phys.\ Rev.\ D {\bf 72}, 026007 (2005)
  [arXiv:hep-th/0505164];
B.A.~Burrington, J.T.~Liu, L.A.~Pando Zayas and D.~Vaman,
  JHEP {\bf 0502}, 022 (2005)
  [arXiv:hep-th/0406207].

\bibitem{spanish} M.~Gomez-Reino, S.~Naculich and H.~Schnitzer,
  Nucl.\ Phys.\ B {\bf 713}, 263 (2005)
  [arXiv:hep-th/0412015].

\end{thebibliography}
\end{document}